\documentclass[12pt]{article}
\usepackage{hyperref}
\usepackage{graphicx}
\usepackage{color}
\usepackage{amssymb} % Symbols for itemise
\definecolor{myblue}{rgb}{0.14,0.11,0.49}
\definecolor{myred}{rgb}{0.74,0.22,0.15}
\definecolor{mygreen}{rgb}{0.05,0.52,0.42}
\definecolor{myyellow}{rgb}{0.96,0.92,0.13}
\definecolor{myorange}{rgb}{1,0.61,0.36}
\definecolor{mypurple}{rgb}{0.71,0.02,1}

\definecolor{htc}{rgb}{1,1,1} % heading text colour
\newcommand{\Couleur}[1]{\textcolor{myblue}{#1}}

\def\be{\begin{equation}}
\def\ee{\end{equation}}
\def\bea{\begin{eqnarray}}
\def\eea{\end{eqnarray}}
\def\bc{\begin{center}}
\def\ec{\end{center}}
\def\bi{\begin{itemize}}
\def\ei{\end{itemize}}

\def\es{\end{slide}}
\def\dd{\mathrm{d}}
\date{}
\title{Representations of the Dirac wave function in a curved spacetime}

\author{
Mayeul Arminjon\,$^{1,2}$ and Frank Reifler\,$^3$\\
$^1$ \small\it CNRS (Section of Theoretical Physics), France.\\
$^2$ \small\it Laboratory ``Soils, Solids, Structures, Risks'' (CNRS, UJF, and G-INP),\\
\small\it BP 53, F-38041 Grenoble cedex 9, France.\\
\small\it $^3$ Lockheed Martin Corporation, MS2 137-205,\\ 
\small\it 199 Borton Landing Road, Moorestown, New Jersey 08057, USA.
} 

\begin{document}
\maketitle

\begin{abstract}
\noindent The Dirac wave function in a curved spacetime is usually defined as a quadruplet of scalar fields. It can alternatively be defined as a four-vector field. We describe these two representations in a common geometrical framework and we prove theorems that relate together the different representations and the different choices of connections. In particular, the standard Dirac equation in a curved spacetime, with any choice of the tetrad field, is equivalent to a particular realization of the Dirac equation for a vector wave function, in the same spacetime. 

\end{abstract}

\section{Introduction}
          
\noindent To our knowledge, the observations of quantum-mechanical properties of matter in the classical gravitational field are the only direct experimental proof of the interplay between gravity and the quantum. Such observations are obtained e.g. by cold neutron interferometry in a plane whose inclination can be varied \cite{COW1975}, or by measuring the transmission of ultra-cold neutrons through a thin horizontal slit \cite{Nesvizhevsky2002}. Since neutrons are spin $\frac{1}{2}$ particles, this motivates work on the curved spacetime Dirac equation. \\

Weyl \cite{Weyl1929b} and Fock \cite{Fock1929b} recognized that, on changing the coordinates, transforming the Dirac wave function under the spin group is not an option in a curved spacetime, or already in a flat spacetime with affine coordinates. They proposed what has become the standard version of the Dirac equation in a curved spacetime \cite{Weyl1929b,Fock1929b}, hereafter referred to as the Dirac-Fock-Weyl equation. As is well known \cite{BrillWheeler1957+Corr, deOliveiraTiomno1962, ChapmanLeiter1977}, the wave function $\Couleur{\psi}\ $ transforms as a quadruplet of complex scalars under a coordinate transformation of the Dirac-Fock-Weyl equation. In contrast, in two recently discovered alternative extensions of the Dirac equation to a curved spacetime, the wave function $\Couleur{\psi}\ $ is a complex four-vector, while the set of the components of the Dirac matrices $\Couleur{\gamma^\mu}$ builds a $\Couleur{(^2 _1)}$ tensor \cite{A39}. Thus, in a curved spacetime, the wave function $\Couleur{\psi}\ $ can be either a quadruplet of complex scalars---this is the Quadruplet Representation of the Dirac field (QRD) to which the Dirac-Fock-Weyl equation belongs. Or, it can be a complex four-vector---this is the Tensor Representation of the Dirac field (TRD) to which the two alternative equations of Ref. \cite{A39} belong. With constant Dirac gamma matrices in a flat spacetime in Cartesian coordinates, all Dirac equations (QRD and TRD) are equivalent to the original Dirac equation with spinor transformation, because: i) The explicit expression of the Dirac equation in a Cartesian coordinate system is the same as with the spinor transformation, and ii) There is no influence of the possible set of constant Dirac matrices \cite{A40}.\\

However, in a general Lorentzian spacetime $\Couleur{(\mathrm{V},g_{\mu\nu})}$ in general coordinates, there is a whole continuum of different possible choices for the field of Dirac matrices, with no possibility of a unique choice \cite{A42,A43,B31}. In our previous studies of the basic quantum mechanics for the standard (Dirac-Fock-Weyl) equation and for the two alternative equations based on TRD, we found that those three equations behave similarly: e.g. the same hermiticity condition of the Hamiltonian applies to the three \cite{A42}, and similar non-uniqueness problems of the Hamiltonian theory occur \cite{A43,B31}. The similar behaviour which we found for the Dirac-Fock-Weyl equation (with $\Couleur{\psi}\ $ being a 4-scalar) and our alternative equations based on TRD led us to study the {\it relations between the two representations in a curved spacetime:} $\Couleur{\psi}\ $ 4-scalar vs. $\Couleur{\psi}\ $ 4-vector.\\

In the work \cite{A45} summarized in the present conference paper, the two representations are formulated in a common geometrical framework. We study and compare the two possible representations of the Dirac wave function in a curved spacetime. We prove several theorems that relate together the QRD and TRD representations in a general spacetime.  \\

%%%%%%%%%%%%%%%%%%%%%%%%%%%%%%%%%%%%%%%%%%%%%%%%%%%%%%%%%%%%%%%%%%%%%%%%%%%%%%%%
  \section{A common geometrical framework}
%%%%%%%%%%%%%%%%%%%%%%%%%%%%%%%%%%%%%%%%%%%%%%%%%%%%%%%%%%%%%%%%%%%%%%%%%%%%%%%%
\noindent The Dirac-Fock-Weyl equation belongs to the more general ``quadruplet representation of the Dirac field'' (QRD). For both QRD and the tensor representation (TRD), the wave function is living in some complex vector bundle with base $\mathrm{\Couleur{V}}$ (the spacetime manifold), and with dimension 4, denoted $\mathrm{\Couleur{E}}$. In spite of being complex, that vector bundle is {\it very simple} in either case: 
\bi
\item For QRD, the wave function is a complex four-scalar field, that is a section of the trivial bundle $\Couleur{\mathrm{V} \times {\sf C}^4}$, thus
\be\label{E-QRD}
\Couleur{\mathrm{E}=\mathrm{V} \times {\sf C}^4} \qquad \mathrm{for\ QRD}.
\ee
\item Instead, for TRD, the wave function is a complex vector field, that is a section of the complexification of the tangent bundle $\Couleur{\mathrm{T}\mathrm{V}}$, thus 
\be\label{E-TRD}
\Couleur{\mathrm{E=T}_{\sf C}\mathrm{V}} \qquad \mathrm{for\ TRD}.
\ee
\ei
\noindent Selecting a local frame field $\Couleur{X \mapsto (e_a(X))}$ of the relevant vector bundle \Couleur{E}, the wave function $\Couleur{\psi}$ has the local expression (valid in some open subset $\Couleur{\mathrm{U}}$ of the spacetime $\Couleur{\mathrm{V}}$)
\be\label{psi=psi^a e_a}
\Couleur{\psi=\Psi^a\,e_a}.
\ee

\vspace{2mm}
\noindent Then, to define the Dirac matrices $\Couleur{\gamma^\mu}$, we may consider a section $\Couleur{\gamma}$ of the following tensor product bundle: 
\be\label{gamm-bundle}
\Couleur{\mathrm{F}=\mathrm{T}\mathrm{V}	\otimes \mathrm{E} \otimes \mathrm{E}^\circ}, 
\ee
where $\Couleur{\mathrm{E}^\circ}$ is the dual vector bundle of \Couleur{E}. To any such section, and depending on the local chart $\Couleur{\chi: X \mapsto (x^\mu)}$ and on the local frame field $\Couleur{(e_a)}$ on \Couleur{E}, both being assumed defined on some open subset \Couleur{U} of the spacetime \Couleur{V}, we associate a unique local expression, valid in \Couleur{U}:
\be\label{gamma-intrinsic}
\Couleur{\gamma = \gamma^{\mu a } _b\ \partial_\mu \otimes e_a \otimes \theta^b},
\ee
with $\Couleur{(\partial_\mu)}$ the coordinate basis associated with the chart $\Couleur{\chi}$ and $\Couleur{(\theta^b)}$ the dual frame of the selected frame field $\Couleur{(e_a)}$. The Dirac matrices $\Couleur{\gamma^\mu}$ are defined to be the matrices with components
\be\label{gamma^mu from gamma intrinsic}
\Couleur{\left(\gamma^\mu\right)^a _{\quad b} \equiv \gamma^{\mu a } _{ b}}, 
\ee
thus they are defined locally and do depend on the local chart and the local frame field. They are of course assigned to satisfy the anticommutation relation 
\be \label{Clifford}
\Couleur{\gamma ^\mu \gamma ^\nu + \gamma ^\nu \gamma ^\mu = 2g^{\mu \nu}\,{\bf 1}_4, \quad \mu ,\nu \in \{0,...,3\} \quad ({\bf 1}_4\equiv \mathrm{diag}(1,1,1,1))}.
\ee
A global section $\Couleur{\gamma }$ of \Couleur{F}, such that (\ref{Clifford}) be thus satisfied in any local coordinates $\Couleur{(x^\mu)}$ and in any local frame field $\Couleur{(e_a)}$, will be called ``an intrinsic field of Dirac matrices''. The existence of at least one intrinsic field of Dirac matrices is guaranteed if the spacetime is four-dimensional, noncompact, and admits a spinor structure. In that case, indeed, there exists an orthonormal global tetrad field \cite{Geroch1968}. Using this fact, it is easy to show that one may define a global section of \Couleur{F}, such that the associated Dirac matrices (\ref{gamma^mu from gamma intrinsic}) satisfy Eq. (\ref{Clifford}) \cite{A45}. This applies to both representations, i.e., whether the wave function lives in the trivial bundle $\Couleur{\mathrm{V} \times {\sf C}^4}$ or in the complex tangent bundle $\Couleur{\mathrm{T}_{\sf C}\mathrm{V}}$.\\

\vspace{5mm}
Select a representation, and consider a $\Couleur{\gamma}$ field for which the anticommmutation relation (\ref{Clifford}) is satisfied, and a connection, say $\Couleur{D}$, on the relevant bundle \Couleur{E}, Eq. (\ref{E-QRD}) or Eq. (\ref{E-TRD}). By the definition of a connection on a vector bundle with base \Couleur{V} \cite{ChernChenLam1999}, it associates to any section $\Couleur{\psi} $ of the bundle \Couleur{E}, a section $\Couleur{D \psi }$ of the tensor product bundle $\Couleur{\mathrm{T V^o\otimes E}}$. (Here $\Couleur{\mathrm{T V^o}}$ is the dual bundle of \Couleur{TV}, hence is the cotangent bundle.) The corresponding Dirac equation in the curved spacetime $\Couleur{(\mathrm{V},g_{\mu\nu})}$ may be written as:
\be\label{Dirac-general-intrinsic}
\Couleur{ \gamma : D\psi =-im\psi},
\ee
where $\Couleur{\gamma : D\psi}$ is the doubly-contracted product of these sections of $\Couleur{\mathrm{T}\mathrm{V}	\otimes \mathrm{E} \otimes \mathrm{E}^\circ}$ and $\Couleur{\mathrm{T V^o\otimes E}}$, corresponding to the two dual pairs $\Couleur{(\mathrm{T V},\mathrm{T V^o})}$ and $\Couleur{(\mathrm{E^o},\mathrm{E})}$ appearing in their respective component spaces. Using a local coordinate system $\Couleur{(x^\mu)}$ and a local frame field $\Couleur{(e_a)}$, $\Couleur{D\psi }$ decomposes as
\be\label{Dpsi}
\Couleur{D\psi=D_\mu \Psi^b\ \dd x ^\mu \otimes e_b},
\ee
with $\Couleur{(\dd x ^\mu)}$ the dual frame of the coordinate frame $\Couleur{(\partial _\mu)}$. The local expression of the Dirac equation (\ref{Dirac-general-intrinsic}) is accordingly:
\be\label{Dirac-local expression}
\Couleur{\gamma^{\mu a } _b D_\mu \Psi^b =-im \Psi^a}.
\ee
This local expression is manifestly covariant under a change of either the coordinate system or the frame field. The connection $\Couleur{D}$ on \Couleur{E} can be given explicitly by providing the expression of the connection matrices $\Couleur{\Gamma_\mu=\left((\Gamma _\mu)^b_{\ \ a}\right)}$, defined by [cf. Eq. (\ref{Dpsi})]:
\be\label{De_a}
\Couleur{D e_a=(\Gamma _\mu)^b_{\ \ a}\ \dd x ^\mu \otimes e_b}.
\ee
The explicit expression of $\Couleur{D \psi }$ is then from (\ref{psi=psi^a e_a}):
\be\label{Dpsi-explicit}
\Couleur{D_\mu \Psi^b \equiv \frac{ \partial \Psi^b}{\partial x^\mu } + (\Gamma _\mu)^b_{\ \ a}\, \Psi^a}.
\ee

\vspace{6mm}
\noindent The link with the more classical writing of the Dirac equation: 
\be\label{Dirac-general-matrix}
\Couleur{ \gamma ^\mu  D_\mu\Psi=-im\Psi}
\ee
is got by defining \Couleur{$\Psi$} and \Couleur{$D_\mu\Psi$} for each $\Couleur{\mu=0,... , 3}$ as column vectors with the following components: 
\be\label{Psi}
\Couleur{\Psi \equiv \left(\Psi^b \right)}
\ee 
and
\be\label{D_mu Psi}
\Couleur{D_\mu \Psi \equiv \left(D_\mu \Psi^b \right)}. 
\ee 
By (\ref{gamma^mu from gamma intrinsic}) and (\ref{D_mu Psi}), the classical writing (\ref{Dirac-general-matrix}) is equivalent to the manifestly-covariant local expression (\ref{Dirac-local expression}). Hence, Eq. (\ref{Dirac-general-matrix}) also is covariant under changes of the coordinate chart and the frame field as well.  \\

\vspace{5mm}
The definitions (\ref{psi=psi^a e_a}) and (\ref{gamma^mu from gamma intrinsic}) give rise to two different transformation behaviours, depending on whether QRD and TRD is considered. \\

For QRD ($\mathrm{\Couleur{E}}= \Couleur{\mathrm{V} \times {\sf C}^4}$), thus including for the standard, Dirac-Fock-Weyl equation, the canonical basis of $\Couleur{{\sf C}^4}$ is a preferred frame field on $\Couleur{\mathrm{E}}$. The wave function $\Couleur{\psi}$ behaves as a scalar under a coordinate change, i.e., it remains {\it invariant}, while the Dirac matrices transform as a four-vector:
\be\label{gamma-vector}
\Couleur{\gamma'^\mu =L^\mu_{\ \nu}\, \gamma^\nu, \qquad L^\mu_{\ \nu} \equiv \frac{\partial x'^\mu }{\partial x^\nu }}.
\ee

\vspace{3mm}
For TRD ($\mathrm{\Couleur{E} = \Couleur{T}}_{\sf \Couleur{C}} \Couleur{\mathrm{V}}$), the frame field on $\Couleur{\mathrm{E}}$ can be taken to be the coordinate basis $\Couleur{(\partial_\mu)}$. In that case, on changing the coordinate chart, $\Couleur{\psi}$ behaves as an usual four-vector:
\be\label{psi'-TRD-coordinate basis}
\Couleur{\Psi'^\mu=L^\mu_{\ \nu} \Psi^\nu},
\ee
and the Dirac matrices transform as an usual $\Couleur{(^2 _1)}$ tensor \cite{A39,A37}:
\be \label{gamma-(^2_1)tensor}
\Couleur{ \gamma'^{\mu \rho} _\nu = L^\mu_{\ \sigma }\,L^\rho _{\ \tau }\,\left(L^{-1}\right)^\chi_{\ \nu  }\gamma^{\sigma \tau } _\chi }.
\ee

\vspace{3mm}
\noindent The anticommutation relation (\ref{Clifford}) is covariant under a change of chart, for either of the two transformation modes (\ref{gamma-vector}) and (\ref{gamma-(^2_1)tensor}), thus including for TRD \cite{A37}. It is also covariant under a change of the local frame field $\Couleur{(e_a)}$ on \Couleur{E}. \\

%%%%%%%%%%%%%%%%%%%%%%%%%%%%%%%%%%%%%%%%%%%%%%%%%%%%%%%%%%%%%%%%%%%%%%%%%%%%%%%%
  \section{Four classes of Dirac equations}
%%%%%%%%%%%%%%%%%%%%%%%%%%%%%%%%%%%%%%%%%%%%%%%%%%%%%%%%%%%%%%%%%%%%%%%%%%%%%%%%
\vspace{2mm}

 1) The standard, Dirac-Fock-Weyl equation, which is a QRD equation, obtains when one assumes simultaneously \cite{BrillWheeler1957+Corr,ChapmanLeiter1977} that:\\

\bi

\item The field $\Couleur{\gamma}$ is deduced from some real {\it tetrad field}:
\be \label{flat-deformed}
\Couleur{\gamma ^\mu(X) = a^\mu_{\ \,\alpha}(X)  \ \gamma ^{ \natural \alpha}}.
\ee
Here $\Couleur{(\gamma ^{\natural \alpha })}$ is a fixed set of constant Dirac matrices, i.e., a constant solution of Eq. (\ref{Clifford}) with $\Couleur{g^{\mu\nu}= \eta^{\mu\nu} \equiv \mathrm{diag}(1,-1,-1,-1)}$. The $\Couleur{a^\mu_{\ \,\alpha}(X)}$ 's are the components in the coordinate basis of a (real) ``tetrad field'' $\Couleur{(u_\alpha)}$, i.e., a frame field on the (real) tangent bundle \Couleur{TV}: $\Couleur{u_\alpha=a^\mu_{\ \,\alpha}\, \partial_\mu}$. The tetrad field $\Couleur{(u_\alpha)}$ is assigned to satisfy the orthonormality condition:
\be\label{orthonormal tetrad}
\Couleur{g_{\alpha \beta} \equiv  g(u_\alpha ,u_\beta )\equiv g_{\mu \nu }\,a^\mu _{\ \,\alpha} \,a^\nu _{\ \,\beta } =\eta _{\alpha \beta }},
\ee
which ensures that the field of matrices (\ref{flat-deformed}) satisfies the anticommutation relation (\ref{Clifford}). \\

\item The connection $\Couleur{D}$ on $\ \Couleur{\mathrm{V} \times {\sf C}^4}\ $ depends on $\Couleur{\gamma}$ in such a way that $\Couleur{D\gamma=0}$.
\ei
\vspace{2mm}
The Dirac-Fock-Weyl equation can then be written as Eq. (\ref{Dirac-general-matrix}).

\vspace{9mm}
\noindent 2) The QRD--0 equations fix the connection $\Couleur{D}$ by assuming that 
 \be\label{QRD-0}
 \Couleur{D \, E_a=0},
 \ee
where $\Couleur{(E_a)}$ is the canonical basis  of $\Couleur{\mathrm{V} \times {\sf C}^4}$. Therefore, the connection matrices $\Couleur{\Gamma_\mu}$ in Eq. (\ref{De_a}) are zero when the frame field $\Couleur{(e_a)}$ is taken to be $\Couleur{(E_a)}$. It follows that the covariant derivatives (\ref{Dpsi-explicit}) entering the Dirac equation (\ref{Dirac-general-matrix}) reduce to partial derivatives:
\be\label{D_mu = drond_mu}
\Couleur{D_\mu \Psi = \partial_\mu\Psi}
\ee
for the QRD--0 version, when the frame field is taken to be $\Couleur{(E_a)}$.\\
 
\vspace{5mm}

\noindent 3) The TRD--0 equations fix the connection $\Couleur{D}$ by assuming that 
\be\label{TRD-0}
\Couleur{D \, u_a=0},
\ee
 where $\Couleur{(u_a)}$ is some global orthonormal (and generally complex) frame field on $\Couleur{\mathrm{T}}_{\sf \Couleur{C}} \Couleur{\mathrm{V}}$. [For the spacetime that we consider, there does exist at least one global orthonormal real tetrad field, say $\Couleur{(u_\alpha)}$, on the real tangent bundle $\Couleur{\mathrm{TV}}$: see after Eq. (\ref{Clifford}). The induced frame field $\Couleur{(u_a)}$ on the complex tangent bundle $\Couleur{\mathrm{T}_{\sf C} \mathrm{V}}$, i.e., $\Couleur{u_a\equiv \delta^\alpha_a u_\alpha}$, is then a global orthonormal frame field on $\Couleur{\mathrm{T}_{\sf C} \mathrm{V}}$.]\\
 
 \vspace{5mm}

\noindent 4) The TRD--1 equations assume the Levi-Civita connection, extended from $\Couleur{\mathrm{T}}\Couleur{\mathrm{V}}$ to $\Couleur{\mathrm{T}}_{\sf \Couleur{C}} \Couleur{\mathrm{V}}$.\\

\vspace{5mm}

For the three classes (2) to (4), the Dirac equation is written with an additional term \cite{A42,A43} as compared with Eq. (\ref{Dirac-general-matrix}):
\be\label{Dirac-general-modified}
\Couleur{\mathcal{D}(\gamma,\mathcal{A},D)\Psi\equiv \gamma ^\mu D_\mu\Psi+\frac{1}{2}A^{-1}(D_\mu B^\mu )\Psi =-im \Psi},
\ee
where 
\be\label{B^mu}
\Couleur{B^\mu \equiv A\gamma ^\mu}.
\ee
In Eq. (\ref{Dirac-general-modified}), $\Couleur{\mathcal{A}}$ is a Hermitian metric with respect to which the Dirac matrices $\Couleur{\gamma^\mu}$ are Hermitian \cite{A40,A45,LawsonMichelson1989}, and $\Couleur{A}$ is the matrix of $\Couleur{\mathcal{A}}$: given a local frame field \Couleur{$(e_a)$} on the bundle \Couleur{E}, with dual frame field \Couleur{$(\theta^a)$} on \Couleur{E$^\circ$}, the local expression of $\Couleur{\mathcal{A}}$ is \cite{A45}
\be
\Couleur{\mathcal{A} = A_{ab} \left(\theta^a\right)^* \otimes\theta^b}
\ee
(where \Couleur{$\left(\theta^a\right)^*$} is the complex conjugate of the one-form \Couleur{$\theta^a$}). This expression is determined by the data of the regular Hermitian \Couleur{$4 \times 4$} matrix field $\Couleur{A\equiv (A_{ab})}$, called the field of the ``hermitizing matrix''  \cite{A40,A42,Pauli1936}. The matrix field \Couleur{$B^\mu$}, Eq. (\ref{B^mu}), is made of all the components at fixed \Couleur{$\mu$} of a tensor field \Couleur{$\mathcal{B}$}, obtained  by contraction:
\be 
\Couleur{\mathcal{B} \equiv \mathcal{A}.\gamma, \quad B^{\mu}_{ab} \equiv A_{ac}\,\gamma^{\mu c}_b}.
\ee
Thus, the matrix $\Couleur{D_\mu B^\mu }$ in Eq. (\ref{Dirac-general-modified}) is made of all components of a second-order tensor field: \Couleur{$D_\mu B^\mu \equiv \left(D_\mu B^{\mu}_{ab}\right)$}. Like Eq. (\ref{Dirac-general-matrix}), Eq. (\ref{Dirac-general-modified}) is covariant under a change of either the coordinate system \Couleur{$(x^\mu)$} or the frame field \Couleur{$(e_a)$}.\\

A direct way of explaining the occurrence of this additional term on the l.h.s. of Eq. (\ref{Dirac-general-modified}) is the following one \cite{A43}: For a general field of Dirac matrices $\Couleur{\gamma^\mu}$, the standard ``Dirac Lagrangian'' which applies to the Dirac-Fock-Weyl equation has to be extended to include the field $\Couleur{A}$ in the exact place of the constant Dirac matrix $\Couleur{\gamma ^{ \natural 0}}$. For the Dirac-Fock-Weyl equation, we have always $\Couleur{D_\mu B^\mu =0}$, hence the additional term in Eq. (\ref{Dirac-general-modified}) vanishes \cite{A42}.\\

The QRD--0 equations, as well as the TRD--0 equations, are new. The TRD--1 equations (together with another class of equations, ``TRD--2'', that is based on a connection deduced from an assumed preferred reference frame) were introduced in Ref. \cite{A39}.\\

\vspace{2mm}
\noindent Note that any two tetrad fields lead to two equivalent Dirac-Fock-Weyl equations \cite{BrillWheeler1957+Corr}, except for some non-trivial topologies \cite{Isham1978}. Thus, the Dirac-Fock-Weyl equation (1) is normally a unique equation, not a class of equations.\\

In contrast, none of the three classes (2) to (4) does correspond to a unique equation. For each of the classes (2) to (4), the connection $\Couleur{D}$ is fixed, but the field $\Couleur{\gamma}$ is restricted only by the anticommutation relation (\ref{Clifford}). In general, two fields $\Couleur{\gamma \ne \gamma'}$ give inequivalent Dirac equations. This is the price to pay for the simplicity of the definition of these equations and for their greater generality, as compared with the Dirac-Fock-Weyl equation. However, if we consider not the Dirac equation itself but the Hamiltonian operator which is associated to the Dirac equation in a given reference frame, then we find \cite{A43,B31} that the  Dirac-Fock-Weyl Hamiltonian is non-unique, too---just as is the Hamiltonian operator associated with, for example, the TRD--1 equation [or with any class of Dirac equations among the classes (2) to (4)]. In other words, the fact that the Dirac-Fock-Weyl equation is unique in a ``reasonable'' spacetime does not prevent the corresponding Hamiltonian from being non-unique (in a given reference frame or even in a given coordinate system).\\

%%%%%%%%%%%%%%%%%%%%%%%%%%%%%%%%%%%%%%%%%%%%%%%%%%%%%%%%%%%%%%%%%%%%%%%%%%%%%%%%
  \section{Equivalence theorems between classes}
%%%%%%%%%%%%%%%%%%%%%%%%%%%%%%%%%%%%%%%%%%%%%%%%%%%%%%%%%%%%%%%%%%%%%%%%%%%%%%%%

In the work \cite{A45}, summarized here, we prove technically precise versions of  three theorems, which we state below in more concise terms. (In addition, Theorem 1 in Ref. \cite{A45} is more general than is Theorem 1 below.)\\

\paragraph {Theorem 1.} {\it The QRD--0 and TRD--0 equations are equivalent for given coefficient fields $\Couleur{(\gamma^\mu,A)}$. Moreover, when the QRD--0 equation is a ``normal'' one, i.e., when $\Couleur{D_\mu B^\mu =0}$ in Eq. (\ref{Dirac-general-modified}), then the corresponding TRD--0 equation is also a normal one.} \\

\noindent Here, ``for given coefficient fields $\Couleur{(\gamma^\mu,A)}$'' means: {\bf i}) that the $\Couleur{\gamma}$ field used to write the QRD--0 equation and the $\Couleur{\gamma'}$ field used to write the TRD--0 equation correspond to the same Dirac matrices in any local coordinate system $\Couleur{(x^\mu)}$ defined on an open subset \Couleur{U} of the spacetime \Couleur{V} --- that is, $\Couleur{\gamma^\mu=\gamma'^\mu}$ --- when the frame fields on \Couleur{E} are, respectively: the canonical basis $\Couleur{(E_a)}$ of $\Couleur{\mathrm{V} \times {\sf C}^4}$, for the QRD--0 equation; and the global tetrad field $\Couleur{(u_a)}$ on $\Couleur{\mathrm{T}_{\sf C} \mathrm{V}}$ that appears in Eq. (\ref{TRD-0}), for the TRD--0 equation. {\bf ii}) Moreover, in these respective frame fields, the respective hermitizing matrix fields coincide, $\Couleur{A=A'}$. With any fields $\Couleur{\gamma}$ and $\Couleur{\mathcal{A}}$ valid for QRD, one may indeed easily associate fields $\Couleur{\gamma'}$ and $\Couleur{\mathcal{A'}}$ valid for TRD, and such that the two equalities $\Couleur{\gamma^\mu=\gamma'^\mu}$ and $\Couleur{A=A'}$ be verified in the conditions specified above \cite{A45}. \\

\noindent The proof of Theorem 1 is not difficult: due to Eqs. (\ref{QRD-0}) and (\ref{TRD-0}) respectively, the covariant derivatives entering the local expression (\ref{Dirac-general-modified}) of the Dirac equation reduce to partial derivatives, both for the QRD--0 or the TRD--0 equation, when the frame field on the relevant bundle \Couleur{E} is $\Couleur{(E_a)}$ or $\Couleur{(u_a)}$, respectively. (See Eq. (\ref{D_mu = drond_mu}) for QRD--0.) And since the coefficient fields $\Couleur{(\gamma^\mu,A)}$ are, by hypothesis, the same for both equations, it follows that the local expression (\ref{Dirac-general-modified}) is the same for the QRD--0 equation and the TRD--0 equation. Thus, the local expressions of these two equations, in the arbitrary local coordinate system $(x^\mu)$ which is considered, and in the respective global frame fields $\Couleur{(E_a)}$ or $\Couleur{(u_a)}$, are indeed equivalent. Finally, to say that the Dirac equation is a ``normal'' one, i.e., that $\Couleur{D_\mu B^\mu =0}$ in Eq. (\ref{Dirac-general-modified}), means for both the QRD--0 and the TRD--0 equation that $\Couleur{\partial_\mu B^\mu =0}$. Since the coefficient fields $\Couleur{(\gamma^\mu,A)}$ are, by hypothesis, the same for the QRD--0 and the TRD--0 equation which are considered, it follows that the validity of the condition $\Couleur{D_\mu B^\mu =0}$ for the QRD--0 equation is equivalent to its validity for the TRD--0 equation. \begin{flushright} $\quad \square$ \end{flushright}

\vspace{5mm}
\paragraph {Theorem 2.}\label{Theorem2} {\it Let $\Couleur{\gamma}$ be any ``intrinsic field of Dirac matrices'', let $\Couleur{\mathcal{A}}$ be an associated hermitizing metric, and let $\Couleur{D}$ be any connection on $\Couleur{\mathrm{E}}$. Let $\Couleur{D'}$ be any other connection on $\Couleur{\mathrm{E}}$.\\

Then, there is another ``intrinsic field of Dirac matrices'', $\Couleur{\tilde{\gamma}}$, with an associated hermitizing metric $\Couleur{\tilde{\mathcal{A}}}$, such that the Dirac equation based on $\Couleur{\gamma}$, $\Couleur{\mathcal{A}}$ and $\Couleur{D}$ is equivalent to that based on $\Couleur{\tilde{\gamma}}$, $\Couleur{\tilde{\mathcal{A}}}$ and $\Couleur{D'}$. In particular, any form of the QRD (TRD) equation is equivalent to a QRD--0 (TRD--1) equation.} \\

\noindent The proof of this theorem \cite{A45} is outlined in the next section.\\

\vspace{2mm}
\paragraph {Theorem 3.}\label{Theorem3} {\it The Dirac-Fock-Weyl equation is equivalent to a TRD--1 equation (thus with vector wave function) in the same spacetime.}\\

\noindent Theorem 3 is an easy consequence of Theorems 1 and 2, at least locally. 

\vspace{2mm}

%%%%%%%%%%%%%%%%%%%%%%%%%%%%%%%%%%%%%%%%%%%%%%%%%%%%%%%%%%%%%%%%%%%%%%%%%%%%%%%%
  \section{Theorem 2: outline of the proof}
%%%%%%%%%%%%%%%%%%%%%%%%%%%%%%%%%%%%%%%%%%%%%%%%%%%%%%%%%%%%%%%%%%%%%%%%%%%%%%%%
For given fields $\Couleur{\gamma}$ and $\Couleur{\mathcal{A}}$, the difference between the Dirac operators $\Couleur{\mathcal{D}(\gamma,\mathcal{A},D)}$ [Eq. (\ref{Dirac-general-modified})] and $\Couleur{\mathcal{D}(\gamma,\mathcal{A},D')}$, corresponding with two different connections $\Couleur{D}$ and $\Couleur{D'}$ on $\Couleur{\mathrm{E}}$, is found after some algebra \cite{A45} to depend just on the matrix
\be\label{K}
\Couleur{K \equiv \gamma^\mu K_\mu},
\ee
so that $\Couleur{\mathcal{D}(\gamma,\mathcal{A},D) = \mathcal{D}(\gamma,\mathcal{A},D')}$ if $\Couleur{K=0}$. Here, the $\Couleur{\gamma^\mu}$ 's are the Dirac matrices associated by Eq. (\ref{gamma^mu from gamma intrinsic}) with $\Couleur{\gamma}$, in the local chart $\Couleur{\chi}$ and in the local frame field $\Couleur{(e_a)}$ which are considered; and we define
\be\label{K=Gamma-Gamma'}
\Couleur{K_\mu \equiv \Gamma_\mu - \Gamma'_\mu},
\ee
where $\Couleur{\Gamma_\mu}$ and $\Couleur{\Gamma'_\mu}$ are the connection matrices (\ref{De_a}) of $\Couleur{D}$ and $\Couleur{D'}$ respectively, in that chart and in that frame field.\\

Consider a new field $\Couleur{\tilde{\gamma}}$ in the same spacetime. In the local chart $\Couleur{\chi}$ and in the local frame field $\Couleur{(e_a)}$, the new field $\Couleur{\tilde{\gamma}}$ corresponds to a new field of Dirac matrices: $\Couleur{X \mapsto \tilde{\gamma}^\mu(X)}$. Similarly, the hermitizing metric $\Couleur{\tilde{\mathcal{A}}}$ associated with $\Couleur{\tilde{\gamma}}$ corresponds in the local frame field $\Couleur{(e_a)}$ to a new field $\Couleur{\tilde{A}}$ of the hermitizing matrix. The field $\Couleur{\tilde{\gamma}}$, as well as the associated hermitizing metric $\Couleur{\tilde{\mathcal{A}}}$, are necessarily deduced from $\Couleur{\gamma}$ and $\Couleur{\mathcal{A}}$ by a local similarity transformation
\be\label{S: V -> GL(4,C)}
\Couleur{S: \mathrm{V} \rightarrow {\sf GL(4,C)}}.
\ee
In the local chart $\Couleur{\chi}$ and in the local frame field $\Couleur{(e_a)}$, this writes in the following way \cite{Pauli1936,A40,A42}:
\be
\Couleur{\tilde{\gamma}^\mu(X) = S(X)^{-1}\gamma^\mu(X)S(X)}
\ee
and
\be
\Couleur{\tilde{A}(X) = S(X)^\dagger A(X)S(X)}.
\ee
We know how to change $\Couleur{D}$ for a new connection $\Couleur{\tilde{D}}$ so that the starting Dirac equation $\Couleur{\mathcal{D}(\gamma,\mathcal{A},D)\Psi=0}$ is equivalent to the Dirac equation $\Couleur{\mathcal{D}(\tilde{\gamma},\tilde{\mathcal{A}},\tilde{D}) \tilde{\Psi}=0}$, where we define $\Couleur{\tilde{\Psi}\equiv S^{-1}\Psi}$.
\footnote{\
Namely, we have to define the new connection by the matrices $\Couleur{\tilde{\Gamma }_\mu = S^{-1}\Gamma _\mu S+ S^{-1}(\partial _\mu S)}$ \cite{A45}.
}
As in Eqs. (\ref{K}) and (\ref{K=Gamma-Gamma'}), set $\Couleur{\tilde{K}_\mu \equiv \tilde{\Gamma}_\mu - \Gamma'_\mu }$ and $\Couleur{\tilde{K}\equiv \tilde{\gamma}^\mu \tilde{K}_\mu }$. Thus, if $\Couleur{\tilde{K}=0}$, then the Dirac equation $\Couleur{\mathcal{D}(\tilde{\gamma},\tilde{\mathcal{A}},D')\tilde{\Psi}=0}$ is equivalent to $\Couleur{\mathcal{D}(\tilde{\gamma},\tilde{\mathcal{A}},\tilde{D})\tilde{\Psi}=0}$, hence to the starting Dirac equation $\Couleur{\mathcal{D}(\gamma,\mathcal{A},D)\Psi=0}$.\\

The condition for $\Couleur{\tilde{K} \equiv \tilde{\gamma}^\mu \tilde{K}_\mu =0}$ can then be shown \cite{A45} to be
\footnote{\
In Eq. (\ref{PDE S}), we have $\Couleur{D' _\mu S=\partial_\mu S + \Gamma'_\mu\,S - S\,\Gamma'_\mu}$ \cite{A43}. This can be defined in intrinsic terms \cite{A45}.
}
\be\label{PDE S}
\Couleur{\gamma ^\mu D' _\mu S =-KS}.
\ee      
This is a system of sixteen first-order linear partial differential equations for the sixteen components of $\Couleur{S}$.  It turns out that this system can be rewritten as a symmetric hyperbolic system \cite{A45}. One consequence is that, by assuming analytic coefficients and initial data, the Cauchy-Kovalevskaya theorem ensures that it can always be solved locally. Another consequence \cite{A45} is that, by using a more advanced theorem, the system (\ref{PDE S}) can be solved in certain a priori given (large) domains. \begin{flushright} $\quad \square$ \end{flushright}

\section{Conclusion}

In the present work \cite{A45}, summarized in this conference paper, we described in a common geometrical framework the Quadruplet Representation of the Dirac field (QRD) and the Tensor Representation of the Dirac field (TRD). We proposed a simple form (``QRD--0'') of the Dirac equation in a curved spacetime, in which the covariant derivatives reduce to \hyperref[D_mu = drond_mu]{partial derivatives}. We proved several theorems that relate together the QRD and TRD representations in a general spacetime. In particular, by \hyperref[Theorem2]{Theorem 2}, the Dirac-Fock-Weyl equation is equivalent to a particular case of the simple QRD--0 equation. By \hyperref[Theorem3]{Theorem 3}, the Dirac-Fock-Weyl equation is also equivalent to a particular case of either of the two alternative equations based on TRD. Thus, these alternative Dirac equations in a curved spacetime, in which the wave function is a complex {\it four-vector}, can be seen as {\it extensions} of the standard version of the Dirac equation in a curved spacetime, in which the wave function is a quadruplet of complex scalars.  \\

\vspace{2mm}
%\newpage

\end{document}